\documentstyle[12pt,epsfig]{article}

\newcounter{multieqs}

\def\be{\begin{equation}}
\def\ee{\end{equation}}
\def\beq{\begin{eqnarray}}
\def\eeq{\end{eqnarray}}
\def\si{\sigma}
\def\te{\theta}

\def\D{\Delta}
\def\p{\partial}

\def\b{\beta}

\def\G{\Gamma}
\def\l{\left}
\def\r{\right}
\def\la{\langle}
\def\ra{\rangle}
\def\pp{\hat{p}}
\def\xx{\hat{x}}

\def\L{\Lambda}

\def\eq{\begin{equation}}
\def\eqe{\end{equation}}
\def\eqa{\begin{eqnarray}}
\def\eqae{\end{eqnarray}}

\def\pa{\partial}
\def\noi{\noindent}
\def\non{\nonumber}

\begin{document}
\baselineskip=18pt
\onecolumn
\hphantom{xxxxxxxxxxxxxxxxxxxxxxxxxxxx}
\hspace{4.5in} ITP-SB-98-04

\vglue 3cm

\begin{center}
{\Large {\bf Mode regularization, time slicing, Weyl ordering and phase space 
path integrals for quantum mechanical nonlinear sigma models.}
}
\vspace{1cm}

Fiorenzo Bastianelli\footnote{Also INFN,
 Sezione di Bologna, Italy. E-mail: bastianelli@imoax1.unimo.it} \\[.5cm]
{\em Dipartimento  di Fisica, Universit\`a di Modena \\
via Campi 213/A, I-41100, Modena, Italy} \\[.5cm]
Koenraad Schalm\footnote{E-mail: konrad@insti.physics.sunysb.edu}  
and Peter van Nieuwenhuizen\footnote{Research supported
by National Science Foundation Grant Phy 9722101.

E-mail: vannieu@insti.physics.sunysb.edu}\\[.5cm]
{\em Institute for Theoretical Physics \\
State University of New York at Stony Brook \\
Stony Brook, New York, 11794-3840, USA}\\
\end{center}
\baselineskip=14pt
\vskip .7cm

\centerline{\large{\bf Abstract}}
\vspace{.25in}

A simple, often invoked, regularization scheme of quantum mechanical path 
integrals in curved space is mode regularization:  one expands fields into a Fourier
series, performs calculations with only the first $M$ modes, and at the
end takes the limit $M \rightarrow \infty$.  This simple scheme
does not manifestly preserve reparametrization invariance of the target 
manifold: particular noncovariant terms of order $\hbar^2$
must be added to the action in order to maintain general coordinate invariance.
Regularization by time slicing requires a 
different set of terms of order $\hbar^2$ which can be derived from
Weyl ordering of the Hamiltonian. With these counterterms both schemes give 
the same answers to all orders of loops. As a check we perform the three-loop 
calculation of the trace anomaly in four dimensions in both schemes. We also 
present a diagrammatic proof of Matthews' theorem that phase space and 
configuration space path integrals are equal.

\newpage
\setcounter{footnote}{0}
\baselineskip=14pt
\section{Introduction}
\hspace{.5in}

Quantum mechanical path integrals in curved target spaces have various
applications.
  Two decades ago they were used to quantize collective coordinates of
solitons \cite{one},
and one decade ago to calculate chiral \cite{two} and trace \cite{three,seven}
anomalies in quantum field theories.
In the former case, the path integrals were regularized  by expanding the
fluctuations about the solitons into normal modes, and cutting the sum
over modes off at a maximum energy or at a maximum number of modes \cite{four}.
In the trivial vacuum, on the other hand, two methods have been widely used:
the time discretization method\footnote{Time discretization in the sector
with solitons is complicated
because the canonical momenta (of the non-zero-modes) satisfy the equal-time 
commutation relation 
$[ \pi(x,t), \varphi (y,t) ] =- i \hbar \mbox{$[ \delta (x-y) - 
\phi^{\prime}_{sol} (x)
\phi^{\prime}_{sol}(y)/M_{sol}]$}$ where $\phi_{sol} (x)$ 
is the classical soliton
solution with mass $M_{sol}$.} which has been pioneered by Dirac and Feynman,
and the mode truncation method already described in \cite{FH}.
In the literature, both methods have been used on a par.

In this article  we point out the relation and the differences
between the two methods.
Both the time discretization method and the mode truncation method
are particular regularization schemes which define the path integral.
One expects therefore that the results for physical objects, for example
the transition element $ T = \langle x^\mu_f ,t_f| x^\mu_i, t_i \rangle $,
can be obtained from path integrals whose actions differ at most by finite
local counterterms.
In quantum gauge field theories, one can fix such
ambiguities by requiring that Ward identities  are satisfied.
Similarly,  for theories in curved target space which are general
coordinate invariant one can require
that the transition element be a ``bi-scalar'' (a scalar in $x^\mu_f$ and
$x^\mu_i$), but this
fixes the ambiguities only up to covariant terms, namely a term proportional 
to the scalar curvature $R$.
Only an experiment, for example the measurement of the trace anomaly,
can fix the coefficient of this term. One can also view the Hamiltonian 
operator $\hat H$ as an observable, and by writing
$T= \langle x^\mu_f ,t_f| x^\mu_i, t_i \rangle = $
$ \la x^\mu_f | {\rm exp}(-{\beta\over \hbar}\hat H)| x^\mu_i \ra $ where $\beta = t_f-t_i$ in Euclidean space,
one can fix the ambiguities in $T$ by requiring that it satisfies
the Schroedinger equation with $\hat H$ as Hamiltonian.
If $\hat H$ itself contains no term proportional to $R$, then
the action will contain a term $-\frac{1}{8}R$ as we will show, and $T$
 will have an
$R$ term in the exponent with coefficient $- {1\over 12}$. 
If $\hat H$ contains a term $\alpha R$, the coefficient of $R$ in $T$ will
be $- ( {1\over 12} + \alpha )$.
In general different regularization schemes can lead to different finite
local
counterterms of any order in $\hbar$, 
but finite local counterterms of order $\hbar$
and $\hbar^2$ only are generated in the path integral action
if one considers different
operator orderings in the Hamiltonian.  A Hamiltonian which is a target
space scalar\footnote{The generator for target space
diffeomorphisms $\hat x^\mu \rightarrow \hat x^\mu + \xi^\mu(\hat x)$
is a sum of an orbital part $\hat G^{orb}(\xi) =
{1 \over { 2 i \hbar}} (\pp_\mu \xi^\mu(\xx) + \xi^\mu(\xx) \pp_\mu )$
and a spin part $ \hat G^{spin}(\xi)= (2 \p_\mu \xi^\lambda g_{\lambda \nu}
- \xi^\lambda \p_\lambda g_{\mu\nu}) {\p \over {\p g_{\mu\nu}}}$.  Closure of
the commutator algebra fixes
$\hat{G}^{spin} (\xi)$.
An example of a target space diffeomorphism scalar is~(\ref{ham}),
namely $[\hat G^{orb}(\xi)+ \hat G^{spin}(\xi), \hat H] =0 $,
as one may check by an explicit (tedious) calculation.}
is
\be 
\hat H = \frac{1}{2} \,g^{-{1\over 4}}(\hat x) \pp_\mu g^{1\over 2 }(\hat x)
g^{\mu\nu}(\hat x)\pp_\nu g^{-{1\over 4}}(\hat x)
\label{ham}
\ee
but changing the order
of the operators will in
general destroy the invariance under target space diffeomorphisms.  If the
ambiguities due to using
different regularization schemes correspond to the ambiguities due to
different operator orderings, the terms of order $\hbar^3$ and higher in the
 Hamiltonian should be the same in all cases. This conclusion is corroborated 
by the observation that in these non-linear sigma models, $N$-loop graphs are
 convergent by power counting for $N \geq3$.
  
In the time discretization method, a very clear connection between
$\hat H$ and the transition element exists
if one uses phase space
path integrals (Feynman's  approach) instead of configuration space path
integrals
(Dirac's approach), and rewrites the Hamiltonian in Weyl ordered form.
Then one may replace the Weyl ordered operator $\hat H_W(\hat p,\hat x)$
in the kernel by the corresponding function at the midpoint
(Berezin's theorem)
\be
\int d p\ \la x_1|{\rm exp}(-{\epsilon \over \hbar} \hat H_W) |p \ra
\la p|x_2 \ra \ \  \rightarrow \ \
{\rm exp} \biggl (-{\epsilon \over \hbar} H (p, {x_1 + x_2 \over 2})
\biggr )\int dp \ \la  x_1|p \ra \la p|x_2 \ra .
\ee
The Weyl ordering of $\hat H$ in (1) leads to the following local
finite correction
to the naive Hamiltonian $H= {1\over 2} g^{\mu\nu} p_\mu p_\nu$
to be used  in the path integral
\be
\Delta V_W = {\hbar^2\over 8} ( R + g^{\mu\nu} \Gamma_{\mu\rho}{}^\sigma
\Gamma_{\nu\sigma}{}^\rho),
\label{eggs}
\ee
where $\Gamma_{\mu\nu}{}^\sigma$ are the Christoffel symbols.\footnote{Our 
conventions for the Riemann and Ricci tensor are:
 $R_{\rho \mu \nu}^{\;\;\;\;\;\; \sigma} = \partial_{\rho} 
\Gamma_{\mu\nu}{}^\sigma + \Gamma_{\rho \tau}{}^{\sigma} 
\Gamma_{\mu\nu}{}^\tau - (\rho \leftrightarrow \mu)$; $R_{\rho \nu} = 
R_{\rho \mu \nu}^{\;\;\;\;\;\; \mu}$. Hence, at the linearized level, 
$R_{\mu\nu} = \frac{1}{2} \left(\pa_{\mu} \pa_{\nu} h-\pa_{\mu} h_{\nu} - 
\pa_{\nu} h_{\mu}+ \Box h_{\mu\nu}\right)$ where $h=\eta^{\mu\nu} g_{\mu\nu}$
 and $h_{\mu}=\eta^{\nu\rho} \pa_{\rho} h_{\mu\nu}$.}  
This counterterm has been extensively discussed and used in the 
literature \cite{six}.

Both in the time discretization and in the mode truncation method, one
writes the paths
as $x^\mu(\tau) = x_{bg}^\mu(\tau) + q^\mu(\tau)$
for $-1 \leq \tau  \leq 0$, where
$x_{bg}^\mu(\tau)$ is a solution of the free field equations with
the correct boundary conditions
($x^\mu_{bg}(\tau)= x_f^\mu +\tau (x_f^\mu- x_i^\mu )$)
and one expands the quantum fluctuations
$q^\mu(\tau)$ into a complete set of eigenfunctions of the free field
equations with vanishing boundary values and
integrates over a finite number of coefficients $q^\mu_m$.  The difference
between the time
discretization method and mode regularization is that in the latter one
uses the continuum
action and naive continuum Feynman rules for a finite number of modes and
one chooses a measure for the
integrals over these modes which is usually normalized such that it
reproduces the standard result for
the free particle if the interaction part of the action, $S_{int}$, is set to
 zero. In the time discretization
method, on the other hand,
one uses a discretized action whose discretized Feynman rules are derived
(in explicit form \cite{nine}) from the
Hamiltonian starting point.

Up to this point we have made fairly obvious statements.
Both regularization schemes are well-defined, and this is how they are used in
the literature. However, it is already clear that this
cannot be the whole truth, because with time slicing one has the
nontrivial $\Delta V_W$ in~(\ref{eggs}), whereas with mode truncation such
a $\Delta V$ seems at first to be absent. The precise way in which also mode 
truncation leads to a
$\Delta V$ will be presented in section~\ref{II}. We will see that one needs
the following counterterm for mode regularization
\be
\Delta V_{MR} = {\hbar^2\over 8} \biggl (
R  -{1\over 3} g^{\mu\nu} g^{\alpha \beta} g_{\gamma\delta}
\Gamma_{\mu\alpha}{}^\gamma \Gamma_{\nu\beta}{}^\delta \biggr).
\ee
As a check we shall obtain the correct trace anomaly in four dimensions from 
a three-loop calculation. In Riemann normal coordinates, the Christoffel 
symbols $\Gamma_{\mu\nu}{}^{\rho}$ in $\Delta V$ vanish at the origin of the
coordinate system, but at the three-loop level one finds a contribution from 
$\Delta V$ by expanding each 
Christoffel symbol into $q^{\tau}\pa_{\tau}\Gamma_{\mu\nu}{}^{\rho}$ and 
contracting the two quantum fields. Since 
$\pa_{\tau}\Gamma_{\mu\nu}{}^{\rho}$ contains a part proportional to the 
Riemann tensor, it is clear that $\Delta V$ yields a nonvanishing
 contribution. Hence dropping $\Delta V$ in Riemann normal coordinates is 
incorrect.

If one eliminates the momenta by integrating over them in the
time discretized path
integral, one obtains $N$ factors $({\rm det} g^{\mu\nu})^{-{1\over2}}$
at the $N$ midpoint coordinates $\frac{1}{2} (q^\mu_{k+1}  + q^\mu_k)$.
Exponentiating these in the
familiar Faddeev-Popov way one obtains what we have called ``Lee-Yang
ghosts'',
namely commuting $a^\mu$ and anticommuting $b^\mu$ and $c^\mu$ ghost fields
 \cite{seven}.
Phase space path integrals are free from ambiguities because they are
finite. Clearly, upon transition from phase space to configuration space,
ambiguities are created; technically this is due to the fact that the momenta
are replaced by the $\dot q\,$'s and ghosts, each of which introduces
divergences and hence ambiguities into the theory.

In the phase space approach, the vertices are different (for example
${1\over 2} p_\mu g^{\mu\nu}(x) p_\nu$ instead of
${1\over 2} \dot x^\mu g_{\mu\nu}(x) \dot x^\nu$), as well as the propagators 
($\la p_\mu p_\nu \ra$ is not proportional to
$g_{\mu\alpha} g_{\nu\beta} \la \dot q^\alpha  \dot q^\beta \ra $),
but the transition element should be the same
(Matthews' theorem). In section~\ref{III} we present a graphical proof.

In section~\ref{IV} we draw conclusions and we show that the coefficient of 
the $R$ term in $\Delta V$ is scheme independent and equals $\frac{1}{8}$.

\section{Mode regularization}
\label{II}
\hspace{.5in}

We now describe how one can define mode regularized path integrals
in curved space. Ideally, one would like to derive mode regularization
from first principles, i.e., starting from the transition amplitude
defined as a matrix element of the evolution operator acting in the
Hilbert space of physical states:
$ \la x^\mu_f | {\rm exp}(-{\beta\over \hbar}\hat H)| x^\mu_i \ra $.
However, this derivation looks quite difficult, so that we prefer to take
a more pragmatic course of action, and attempt a direct definition of
the mode regularized path integral.
This definition will be supplemented by certain
consistency requirements which we specify later on.

The transition amplitude can formally be written  as follows
\beq
&& \langle x^\mu_f,t_f | x^\mu_i,t_i \rangle  =
\int_{_{x(-1) = x_i}}^{^{x(0)=x_f}} 
\tilde {\cal D} x\ \exp \biggl [ - { 1\over {\beta \hbar }} S \biggr ]; 
\label{pi2}\\
&& S = \int_{-1}^{0} \, d\tau \
{1\over 2} g_{\mu\nu}(x) (\dot x^\mu \dot x^\nu + a^\mu a^\nu +
b^\mu c^\nu),  \nonumber \\
&& \tilde {\cal D} x
= \prod_{{-1}< \tau < 0}  d^D x(\tau) d^D a(\tau) d^D b(\tau) d^D c(\tau).
\nonumber
\eeq
We have shifted and rescaled the time parameter $t-t_f=\beta \tau$, and 
since all terms in the action only depend
on $\beta \hbar$, we set $\hbar=1$ from now on.  Note that $\beta$ then 
counts the
number of loops.  We will
evaluate the path integral in a perturbative expansion in $\beta$
and in  the coordinate displacements $\xi^{\mu}$ about the final point:
$\xi^\mu = x^\mu_i - x^\mu_f$.
Formally the path integral is a scalar since the action is a scalar 
and the ghost fields make up a scalar measure on the space of
paths.\footnote{The factors $\prod (\det g(x_i))^{1/2}  d x_i^{\mu}$ in the
 discretized path integral are target space scalars and exponentiating them 
leads to ghosts.} Mode regularization will destroy this formal covariance.
However, we will see that
covariance can be recovered by adding a suitable noncovariant counterterm
$\Delta V$ to the action $S$.

We start parametrizing
\be
x^\mu(\tau) = x_{bg}^\mu(\tau) + q^\mu(\tau),
\ee
where $x_{bg}^\mu(\tau) $ is a background trajectory and
$q^\mu(\tau)$ the quantum fluctuations.
The background trajectory is taken to satisfy the free field equations of 
motion and is a function linear in $\tau$
connecting $x^\mu_i$ to $x^\mu_f$ in the chosen coordinate system, thus
enforcing the boundary conditions
\be
 x_{bg}^\mu(\tau) = x_f^\mu  - \xi^\mu \tau; \;\;\;\; \xi^{\mu} = x^{\mu}_i - 
x^{\mu}_f .
\ee
Then the quantum fields $q^\mu (\tau)$ should vanish at the time boundaries
and can
be expanded into sines. For the Lee-Yang ghosts we use the same Fourier 
expansion; this may be considered as part of our definition of mode 
regularization.\footnote{Another argument to justify that the ghosts should 
be expanded into sines is that the classical solutions of their field 
equations are $a^{\mu}=b^{\mu}=c^{\mu}=0$, and that the quantum fluctuations 
do not modify the boundary conditions of the classical solutions.
 In~\cite{nine} it was shown that the results for the transition amplitude
 do not change if one uses cosines for the ghosts.} Hence

\beq
&& q^\mu(\tau) =\sum_{m=1}^{\infty} q^\mu_m \sin (\pi  m \tau) ;\ \ \
a^\mu(\tau) = \sum_{m=1}^{\infty} a^\mu_m \sin (\pi  m \tau); \nonumber\\
&& b^\mu(\tau) = \sum_{m=1}^{\infty} b^\mu_m \sin (\pi  m \tau) ;\ \ \
c^\mu(\tau) = \sum_{m=1}^{\infty} c^\mu_m \sin (\pi  m \tau).
\eeq

At this point the formal measure $\tilde {\cal D}x $
can be defined in  terms of integration over the Fourier coefficients
\be
\tilde {\cal D}x =
{\cal D}q {\cal D}a {\cal D}b {\cal D}c = 
A \prod_{m=1}^{\infty}
\prod_{\mu=1}^D 
d q^\mu_m  d a^\mu_m  d b^\mu_m d c^\mu_m, \label{mes}
\ee
which fixes the path integral for a free particle up to the constant $A$
\be
\int \tilde {\cal D} x
\ \exp \biggl [ - { 1\over \beta} Q \biggr ] = A ; \ \ \ \
Q = \int_{-1}^{0}  d\tau\
{1\over 2} \delta_{\mu\nu} ( \dot q^\mu  \dot q^\nu
+ a^\mu a^\nu + b^\mu c^\nu) .
\label{mes2}
\ee
The constant $A$ will be fixed later on from a consistency
requirement. Note that limiting the integration over the number of modes to a 
finite number $M$ gives a natural
regularization of the path integral.
This regularization resolves the
ambiguities that show up in the continuum limit.

Now we expand the action about the final point $x^\mu_f$ and obtain
\be
S = S_2 + S_{int} = S_2 + S_3 + S_4 + \dots
\ee
where
\beq
S_2 &=& \int_{-1}^{0}  d\tau\  {1\over 2}
g_{\mu\nu}(\xi^\mu \xi^\nu +
\dot q^\mu \dot q^\nu + a^\mu a^\nu +b^\mu c^\nu),
\nonumber\\
S_3 &=& \int_{-1}^{0}  d\tau\ {1\over 2}
\partial_\alpha  g_{\mu\nu} (q^\alpha -\xi^\alpha \tau)
(\xi^\mu \xi^\nu + \dot q^\mu \dot q^\nu + a^\mu a^\nu +b^\mu c^\nu
- 2 \dot q^\mu  \xi^\nu),
\\
S_4 &=& \int_{-1}^{0}   d\tau \
{1\over 4} \partial_\alpha \partial_\beta g_{\mu\nu}
(q^\alpha q^\beta + \xi^\alpha \xi^\beta \tau^2 - 2 q^\alpha \xi^\beta \tau)
(\xi^\mu \xi^\nu + \dot q^\mu \dot q^\nu + a^\mu a^\nu +b^\mu c^\nu
-2 \dot q^\mu \xi^\nu). \nonumber
\eeq
All geometrical quantities, like $g_{\mu\nu}$ or
$\partial_\alpha  g_{\mu\nu}$, are evaluated at the final point $x^\mu_f$,
but for notational simplicity we do not exhibit this dependence.
$S_2$ is taken as the free part and defines the propagators
while $S_{int}$ gives the vertices as usual.
Therefore, the quantum perturbative expansion reads:
\beq
&& \langle x^\mu_f,t_f | x^\mu_i,t_i \rangle  =
\int (\tilde {\cal D} x)
\ \exp \biggl [ - {1\over \beta} S \biggr ]
= A {\rm e}^{-{1\over 2\beta }g_{\mu\nu}\xi^\mu\xi^\nu}
\langle {\rm e}^{- {1\over \beta} S_{int}} \rangle \nonumber\\
&& = A {\rm e}^{-{1\over {2\beta}} g_{\mu\nu}\xi^\mu \xi^\nu}
\biggl (\langle 1 -
{1\over \beta }S_3 - {1\over \beta } S_4 + {1\over  2\beta^2}
S_3^2 \rangle + O(\beta^{3\over2}) \biggr ). \label{pex}
\eeq
Aiming at a two-loop computation, we have kept
only those terms contributing up to $O(\beta)$, taking
into account that $\xi^\mu \sim
O({\sqrt{\beta}})$, as follows from the exponential appearing in the
last line of~(\ref{pex}).
The propagators that follow from $S_2$ are given by 
\beq
\la q^\mu(\tau) q^\nu(\si)\ra &=&
-\beta g^{\mu\nu}(x_f) \D(\tau,\si)
\nonumber\\
\la a^\mu(\tau) a^\nu(\si)\ra &=&  \beta g^{\mu\nu}(x_f)
{}^{\bullet\bullet\!}\D (\tau, \sigma)
\\
\la b^\mu(\tau) c^\nu(\si)\ra &=& -2\beta g^{\mu\nu}(x_f)
{}^{\bullet\bullet\!}\D (\tau,\sigma)
\nonumber
\eeq
where
$\Delta$ is regulated by the mode cut-off introduced below~(\ref{mes2}):
\be
\Delta (\tau,\sigma) = \sum_{m=1}^{M}
\biggl [ - {2 \over {\pi^2 m^2}} \sin (\pi m \tau)
\sin (\pi m \sigma) \biggr ],  \label{modereg}
\ee
and has as continuum value
\be
\D(\tau,\sigma)=\tau(\si+1)\te(\tau-\si)+\si(\tau+1)\te(\si-\tau)\, .
\ee
(A dot on the left(right) of $\Delta (\tau,\sigma)$ indicates differentiation 
with respect to $\tau(\sigma)$.) Using standard Wick contractions, we computed:
\beq
\langle - {1\over \beta} S_3 \rangle &=& - {1\over \beta} {1\over 4}
\p_\alpha g_{\mu\nu} \xi^\alpha \xi^\mu \xi^\nu,
\\
\langle - {1\over \beta} S_4 \rangle &=&
\p_\alpha \p_\beta g_{\mu\nu} \biggl [ {\beta\over 24}
(g^{\mu\nu} g^{\alpha\beta} -g^{\mu\alpha} g^{\nu\beta})
-{1\over 24}(g^{\mu\nu} \xi^\alpha \xi^\beta +g^{\alpha\beta}
\xi^\mu \xi^\nu - 2 g^{\mu\alpha} \xi^\nu \xi^\beta ) \nonumber \\
&& -{1\over\beta}{1\over 12} \xi^\mu \xi^\nu \xi^\alpha \xi^\beta \biggr],
\label{s4}\\
\langle {1\over 2\beta^2} S_3^2 \rangle &=&
\p_\alpha g_{\mu\nu} \p_\beta g_{\lambda\rho} \biggl [
{\beta \over 96} (g^{\alpha\beta} g^{\mu\nu} g^{\lambda \rho}
- 4 g^{\alpha\rho} g^{\mu\nu} g^{\beta\lambda}
- 6 g^{\alpha\beta} g^{\mu\lambda} g^{\nu\rho}
\nonumber \\ &&
+ 4 g^{\alpha\rho} g^{\beta\mu} g^{\nu\lambda}
+ 4 g^{\alpha \mu} g^{\beta\lambda} g^{\nu\rho} )
\nonumber \\ &&
+{1\over 48} \biggl(g^{\mu\lambda} g^{\nu\rho} \xi^\alpha \xi^\beta
+ 2( g^{\alpha\beta} g^{\mu\lambda} - g^{\alpha\lambda} g^{\mu\beta})
\xi^\nu \xi^\rho
+ (2 g^{\alpha\lambda} g^{\beta\rho}-  g^{\alpha\beta} g^{\lambda\rho})
\xi^\mu \xi^\nu
\nonumber\\ &&
+ (2  g^{\mu\beta} g^{\lambda\rho} - 4 g^{\mu\lambda} g^{\beta\rho})
\xi^\alpha \xi^\nu \biggr )
\nonumber\\ &&
+{1\over \beta} {1\over 96} (g^{\alpha\beta} \xi^\mu \xi^\nu \xi^\lambda
\xi^\rho
- 4 g^{\alpha\lambda} \xi^\mu \xi^\nu \xi^\beta \xi^\rho
+ 4 g^{\mu\lambda} \xi^\alpha \xi^\nu \xi^\beta \xi^\rho )
\nonumber\\ &&
+{1\over \beta^2}{1\over 32} \xi^\alpha \xi^\mu \xi^\nu \xi^\beta
\xi^\lambda \xi^\rho \biggr ] .
\eeq
This completes the calculation of the transition amplitude in the
two-loop approximation using the mode regularized path integral.

At this point we should make contact with other schemes and
test the consistency of our rules.
To do that, we use our transition amplitude to obtain the time evolution of an
arbitrary wave function
\be
\Psi(x_f,t_f) = \int \! d^D \!x_i \sqrt{g(x_i)}
\langle x^\mu_f,t_f | x^\mu_i,t_i \rangle  \Psi(x_i,t_i) .
\label{test}
\ee
We need the factor $\sqrt{g(x_i)}$ because the transition element is formally 
a biscalar as we explained before, but then $\Psi(x_f,t_f)$ is a scalar and 
 hence also $\Psi(x_i,t_i)$, which in turn implies that the measure must be a
 scalar as well. 

Since the transition amplitude~(\ref{pex}) is given in terms of an expansion 
around
the final point $ (x_f,t_f)$, we Taylor expand also
the wave function $ \Psi(x_i,t_i) $ and the measure
$ \sqrt{g(x_i)} $ in  eq.~(\ref{test}) 
about this point
\beq
\Psi(x_i,t_i) &=&
\Psi(x_f,t_f) - \beta \p_t \Psi(x_f,t_f)
+ \xi^\mu \p_\mu \Psi(x_f,t_f) +
{1\over 2} \xi^\mu \xi^\nu \p_\mu \p_\nu \Psi(x_f,t_f) +O(\beta^{\frac{3}{2}})
\nonumber\\
\sqrt{g(x_i)} &=&
\sqrt{g(x_f)} \biggl (1 +\xi^\mu \G_{\mu\alpha}{}^\alpha
 +{1\over 2} \xi^\mu\xi^\nu (\p_\mu \G_{\nu\alpha}{}^\alpha +
\G_{\mu\alpha}{}^\alpha \G_{\nu\beta}{}^\beta) + O(\xi^3)
 \biggr ).
\eeq
We then perform the integration over $d^D x^{\mu}=d^D\xi^{\mu}$ 
in~(\ref{test}) and match the various terms.
The leading term fixes $A$
\be
\Psi = A (2 \pi \beta)^{D\over2} \Psi
\ \ \ \ \ \rightarrow \ \ \ \ \
A = (2 \pi \beta)^{-{D\over2}}, \label{zero}
\ee
and the terms of order $\beta$ give
\be
\beta \biggl [ - \p_t \Psi + {1\over 2} \nabla^2 \Psi + {1\over 8} R \Psi
-{1\over 32}  g^{\mu\nu} g^{\alpha \beta} g^{\gamma\delta}
\p_\mu g_{\alpha \gamma} \p_\nu g_{\beta\delta}\Psi
+{1\over 48}   g^{\mu\nu} g^{\alpha\beta} g^{\gamma\delta}
\p_\mu g_{\alpha\gamma} \p_\beta g_{\nu\delta}
\Psi \biggr ]   = 0 .
\ee
This last equation means that the wave function $\Psi$
satisfies the following Schroedinger equation at the
final point  $(x^\mu_f, t_f)$:
\be
- \p_t \Psi = (H_0 + \Delta V_{eff}) \Psi =
 -{1\over 2} \nabla^2 \Psi +  \Delta V_{eff} \Psi,
\ee
where the effective potential is given by
\beq
\Delta V_{eff} &=&
- {1\over 8} R
+{1\over 32}  g^{\mu\nu} g^{\alpha \beta} g^{\gamma\delta}
\p_\mu g_{\alpha\gamma} \p_\nu g_{\beta\delta}
- {1\over 48}   g^{\mu\nu} g^{\alpha \beta} g^{\gamma\delta}
\p_\mu g_{\alpha\gamma} \p_\beta g_{\nu\delta}
\nonumber \\
&=& -{1\over 8} R  - {1\over 8} g^{\mu\nu} \Gamma_{\mu\alpha}{}^\beta
\Gamma_{\nu\beta}{}^\alpha
+{1\over 24}   g^{\mu\nu} g^{\alpha\beta} g^{\gamma\delta} \p_\mu
g_{\alpha\gamma} \p_\beta g_{\nu\delta}
\label{countmr}
\\
&=& -{1\over 8} R  +{1\over 24} g^{\mu\nu} g^{\alpha \beta} g_{\gamma\delta}
\Gamma_{\mu\alpha}{}^\gamma \Gamma_{\nu\beta}{}^\delta.
\nonumber
\eeq
Clearly, to obtain the ``free'' Hamiltonian $H_0$ from a path integral,
one should subtract the potential $\Delta V_{eff}$,
and thus use in~(\ref{pi2}) the following classical action
\be
-\frac{1}{\beta} S  = \int_{-1}^{0} d\tau\ \biggl (
-{1\over 2 \beta} g_{\mu\nu}(x) \dot x^\mu \dot x^\nu
+ \beta \Delta V_{eff}(x) \biggr ).
\ee
With this counterterm mode regularization gives the same results as the time 
discretization
scheme described in the introduction. Note that~(\ref{countmr}) is different 
from the
counterterm for time discretization. 
The difference is the last term in the second line of~(\ref{countmr}).

Finally, as a non-trivial test, which  was one of our motivations to carry
out the present research, we have computed the trace anomaly for a real 
conformal scalar field in four dimensions
using mode regularization and the newly found counterterm. Anomalies in 
Fujikawa's approach are given by the regulated trace of the Jacobian $J$ of 
the symmetry transformation. 
\be 
An = \lim_{\beta \rightarrow 0} \mbox{Tr}(J \; e^{-\frac{\beta}{\hbar} 
{\cal R}}). 
\label{ann}
\ee
A theory of consistent regulators ${\cal R}$ exists~\cite{acht} and it gives
 us for conformal scalar fields the Hamiltonian~(\ref{ham}) minus an 
improvement potential $\,\hbar^2 \frac{D-2}{8(D-1)}R$. The Jacobian for trace
 anomalies is proportional to unity. In mode regularization~(\ref{ann}) 
translates into the evaluation of the expectation value of the Jacobian with 
respect to the path integral based on the formal transition 
element~(\ref{pi2}) including the just calculated counterterms and the 
improvement term. Taking the trace of~(\ref{pi2}) means first 
putting $x_f=x_i$ which gives that $x_{bg}=x_f$ and 
then integrating over $x_f$ with $d^Dx_f \sqrt{g(x_f)}$.
As the measure of~(\ref{pi2}) includes the
constant $A=(2 \pi \beta)^{-\frac{D}{2}}$, the $\beta$-independent term of
 the r.h.s. of~(\ref{ann}) is given by the $(D/2+1)$-loop term in the path 
integral. Expanding the metric in Riemann normal coordinates we find the 
first nonvanishing contributions at the two-loop level: graphs with the 
topology of the number 8 and the background values of the various potential 
terms. From~(\ref{s4}) with $\xi^{\mu}=0$ one obtains\footnote{In equations 
(\ref{twoloop}), (\ref{three1}), (\ref{eye}), (\ref{glover}) only the 
topology of the graphs is indicated, e.g., the figure 8 in (\ref{twoloop})
 stands for all possible graphs of that shape, including ghost loops.}
\be
\raisebox{-.1in}{
\epsfysize=0.3in
\epsfbox{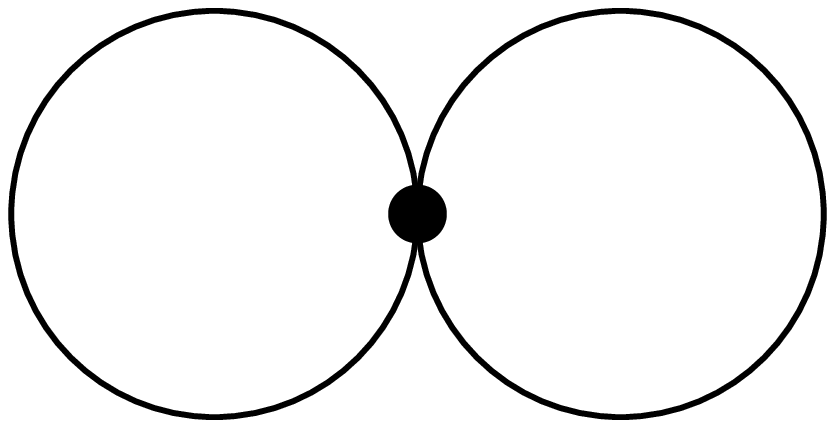}} \hspace{0.3in} + \hspace{0.2in} \bullet \hspace{.2in} = 
(-\beta \hbar)\frac{1}{6}\left(-\frac{1}{4}\right)R + 
(-\beta \hbar)\frac{(1-(D-2)/(D-1))}{8}R.
\label{twoloop}
\ee
The result is the same as obtained from time slicing and yields the correct
 trace anomaly in $D=2$ dimensions.

At the three-loop level we find three regular graphs
\eqa
\raisebox{-.1in}{
\epsfxsize=1.2in
\epsfbox{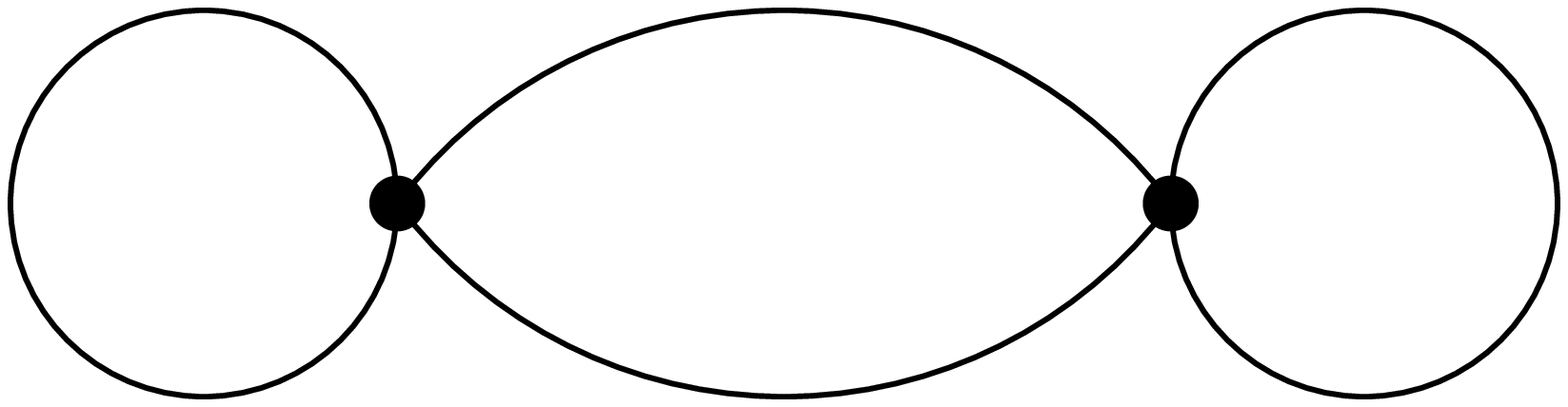}}
\hspace{0.5in}
&=& \hspace{0.5in}\frac{1}{72} (-\beta\hbar)^2 \left(-\frac{1}{6}
 R_{\mu\nu}^2 \right),
\label{three1}
\\[.1in]
\raisebox{-.2in}{
\epsfxsize=0.5in
\epsfbox{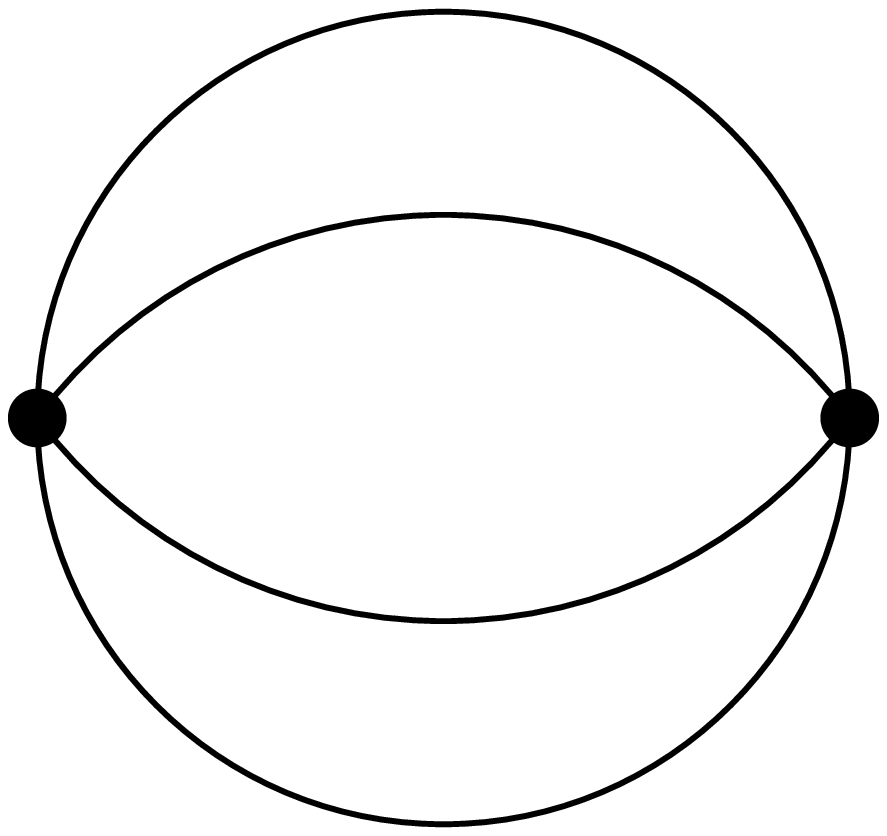}}
\hspace{0.5in}
&=& \hspace{0.5in} \frac{1}{72} (-\beta\hbar)^2 \left(-\frac{1}{6} 
R_{\mu\nu\lambda\rho}^2 \right),
\label{eye}
\\[.2in]
\raisebox{-.25in}{
\epsfxsize=0.5in
\epsfbox{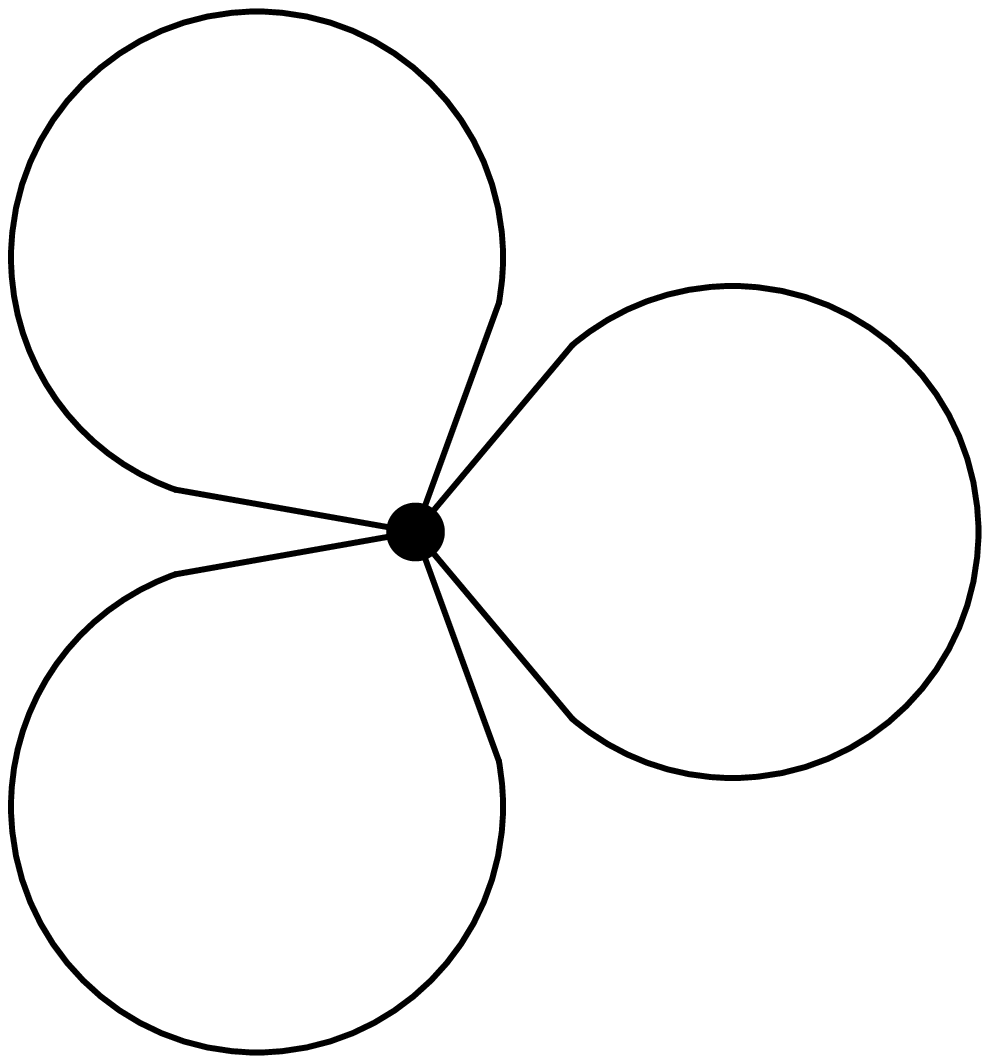}}
\hspace{0.5in}
&=& (-\beta\hbar)^2 \left(\frac{1}{480} \nabla^2 R + \frac{1}{720} 
R_{\mu\nu\lambda\rho}^2 + \frac{1}{1080} R_{\mu\nu}^2 \right),
\label{glover}
\eqae
as well as a graph coming from the potential terms
\be
\raisebox{-.15in}{\epsfysize=0.4in
\epsfbox{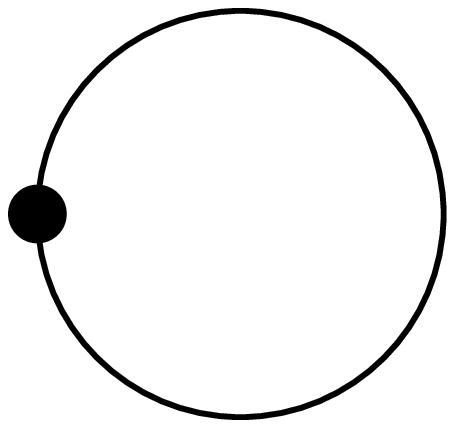}}
\hspace{0.5in}
= \hspace{0.5in} -\frac{1}{6} (\beta\hbar)^2 \left(\frac{1-(D-2)/(D-1)}{16} 
\nabla^2 R - \frac{1}{72} R_{\mu\nu\lambda\rho}^2 \right).
\label{vertex}
\ee

\noi There are only two differences with respect to time slicing: with time 
slicing, the factor $-\frac{1}{6}$ in~(\ref{eye}) becomes $-\frac{1}{4}$, and 
the factor $-\frac{1}{72}$ in~(\ref{vertex}) becomes $-\frac{1}{48}$. These 
two modifications lead to the same final expression. Adding all contributions
 of connected and disconnected graphs we find the correct result   

\be
An(Weyl) (spin \,0 , D=4) = \int \frac{d^4 x}{(2\pi)^2} \sqrt{g(x)} \sigma(x)
 \frac{1}{720} \left(R_{\mu\nu\lambda\rho}^2 - R_{\mu\nu}^2 - \nabla^2 R 
\right).
\label{anom4}
\ee

\noi where $\sigma(x)$ is the arbitrary function appearing in the Jacobian:
$J= \sigma(x) \delta^D(x -y)$.
This shows that mode regularization works after
all.\footnote{ In \cite{seven} the noncovariant part of the counterterm 
(\ref{countmr}) was missed.
In Riemann normal coordinates this term affects only the coefficient
of the $R_{\mu\nu\lambda\rho}^2$ term in the four dimensional trace anomaly, 
which
was erroneously calculated.}
The fact that the right answer is obtained suggests
that no new counterterms are needed in this scheme.
A complete three-loop calculation in arbitrary coordinates
and with a non-vanishing $\xi^\mu$ could be used
to test eq.~(\ref{test}) at order $\beta^2$. As we noted in the introduction,
 however, all three- and higher-loop graphs are power-counting convergent. This
 means that at these orders in $\hbar$ any consistent regularization scheme
 will yield the same answer. 
We therefore claim that with the presently found counterterm,
the mode regularized path integral is consistent to all orders.
This suggests that the difference is indeed due to operator-orderings. It 
would be interesting to justify mode regularization
from first principles.

Before closing this section, we comment on how loop integrations
in Feynman diagrams are done.
Mode cut-off allows one to disentangle ambiguities that appear
in the continuum limit of certain integrals over the $\Delta$'s.
Resorting to the mode regulated expression for $\Delta$,
one can use partial integration and take
boundary terms into account if they are non-vanishing.
Using partial integration repeatedly, one gets
expressions containing only $\Delta$'s, $\Delta^{\bullet}$'s and 
${}^{\bullet}\Delta$'s  which are unambiguous in the continuum limit,
and computes them there.
Useful identities obtained at the regulated level
from~(\ref{modereg}) are
\beq
&& {}^\bullet\! \Delta^\bullet (\tau,\tau)+ {}^{\bullet \bullet} \Delta
(\tau,\tau)
= \p_\tau ({}^\bullet\! \Delta (\tau,\tau)),
\nonumber\\
&& {}^{\bullet \bullet} \Delta (\tau,\sigma) =
\Delta^{\bullet\bullet} (\tau,\sigma).
\eeq
The first identity was used to compute~(\ref{twoloop}). Here is a list of 
integrals that are easily computed
using partial integration\footnote{To obtain $I_2$, use 
$(\Delta^{\bullet})({}^{\bullet}\Delta^{\bullet})=
\frac{1}{2}\pa_{\tau}(\Delta^{\bullet})^2$, partially integrate, 
use ${}^{\bullet\bullet}\Delta = \Delta^{\bullet\bullet}$, and then use
 $(\Delta^{\bullet\bullet})(\Delta^{\bullet})^2=\frac{1}{3}\pa_{\sigma}
(\Delta^{\bullet})^3$. 
To obtain $I_3$, write the integrand as $\Delta({}^{\bullet}\Delta)\frac{1}{2} 
\pa_{\tau} (\Delta^{\bullet})^2$, partially integrate, and use then that 
$\Delta({}^{\bullet\bullet}\Delta) (\Delta^{\bullet})^2 = 
\Delta(\Delta^{\bullet\bullet})(\Delta^{\bullet})^2= \Delta 
\frac{1}{3}\pa_{\tau}(\Delta^{\bullet})^3$. Then partially integrate once
 more.} and whose values are {\em different}
from the time discretization method:
\beq
&&
I_1 = \int d\tau\  \tau \ ({}^\bullet\! \Delta^\bullet +
{}^{\bullet\bullet} \Delta)|_{\sigma=\tau}
= 0,
\nonumber\\
&& I_2 = \int\!\!\int d\tau d\sigma\  ({}^\bullet\! \Delta)\
(\Delta^\bullet)\
({}^\bullet\! \Delta^\bullet)  = - {1\over 12},
\label{bacon}
\\
&& I_3 = \int\!\!\int d\tau d\sigma\
\Delta\ ({}^\bullet\! \Delta)\ (\Delta^\bullet)\
({}^\bullet\! \Delta^\bullet) =
{1\over 180}.
\nonumber
\eeq
Time discretization would give $I_1= -{1\over2}$,
$I_2= -{1\over 6}$ and  $I_3= {7\over 360}$.\footnote{In the time 
discretization scheme \mbox{$\delta(\sigma -\tau)$} is a Kronecker delta 
function and $\theta(0)=\frac{1}{2}$. This leads to a fully consistent scheme 
as shown in \cite{nine}. For example, $\int \!\! \int d\sigma \, d\tau\,
\mbox{$\delta(\sigma-\tau)
$} \mbox{$\theta(\sigma-\tau)$}\theta(\sigma-\tau)  = \int \! \!\int d\sigma
 \, d\tau\,\delta(\sigma-\tau) \theta(\sigma-\tau) \theta(\tau-\sigma) 
=\frac{1}{4}$ whereas mode regularization would give $\frac{1}{3}$ and
 $\frac{1}{2}-\frac{1}{3}=\frac{1}{6}$ respectively.  Using ${}^{\bullet} 
\Delta(\tau,\sigma) = \sigma + \theta(\tau-\sigma)$, ${}^{\bullet} 
\Delta^{\bullet}(\tau,\sigma)=1-\delta(\tau-\sigma)$ and ${}^{\bullet\bullet}
 \Delta = \delta(\tau-\sigma)$, the results below~(\ref{bacon}) follow.\label{tmdsc}}

The non-vanishing value of $I_1$  in time discretization is needed
to compensate the explicit factors of $g^{-{1/4}}(x_i)$ appearing
in the measure, see~(\ref{pi1}), and together with $I_2$ it is responsible
 for the
different counterterms required in the two regularization schemes.
Finally, $I_3$ leads to the different values for the coefficient of
 $R^2_{\mu\nu\lambda\rho}$ in eq.~(\ref{eye}) in the two schemes. The 
counterterms with two Christoffel symbols are also different in both schemes 
but the final result for the transition element (and hence the trace anomaly)
 is the same. This confirms our approach to mode regularization. 

\section{Phase space path integrals in curved space and Matthews' theorem}
\label{III}
\hspace{.5in}

In the phase space approach to path integrals in curved space the
vertices and propagators are different from those used
in the configuration space approach.
For example, the leading terms in the action are
${1\over 2} p_\mu g^{\mu\nu}(x) p_\nu$
and
${1\over 2} \dot x^\mu g_{\mu\nu}(x) \dot x^\nu$,
respectively, and the
$\la\dot q^\mu(\tau) \dot q^\nu(\sigma) \ra$
propagator contains a Dirac delta function but the
$\la p_\mu(\tau) p_\nu(\sigma)\ra$
propagator is completely regular.
Although the vertices and propagators differ, the result for the transition
element should be the same (Matthews' theorem\footnote{Matthews' theorem holds
for connected graphs and originally only applied to nonderivative 
interactions.})~\cite{elf}, and the way this comes
about is due to a new kind of ghosts \cite{three,seven}.
The difference  proportional to $\delta(\tau - \sigma)$
between the $pp$ and $\dot q \dot q$ propagators turns out
to be the  propagator of those  ghosts which we have called
Lee-Yang ghosts after Lee and Yang, who where the first to point
out that in the deformed harmonic oscillator (using arbitrary
coordinates in one dimension) one needs extra $\delta(0)$
terms in the action or Hamiltonian \cite{tien}. The modern way to deal with
these ill-defined objects $\delta(0)$ is to introduce new ghosts,
just as in the familiar case of Faddeev-Popov ghosts for gauge
theories.  From the formal definition of the path integral, it is clear
that Matthews' theorem should hold~\cite{negen}, but as always it is desirable
 to check this by
a nontrivial calculation.
Moreover, such a calculation in turn may reveal a relation between
the  Feynman graphs in both approaches which can lead to
a diagrammatic proof of Matthews' theorem.
With this in mind we calculate the two-loop correction
to the transition element
$ \langle x^\mu_f , 0| x^\mu_i, -\beta \rangle $,
both using the phase space approach and the configuration space
approach to path integrals.

Using time discretization (see footnote~\ref{tmdsc}), the transition element is given in both phase- and 
configuration-space by
\be
 \la x_f|\exp (-{\b\over\hbar}\hat H)|x_i\ra=
\l[ {g(x_f)\over g(x_i)} \r]^{1/4}
{1\over (2\pi\b\hbar)^{D/2}} \la \exp (-{1\over \hbar}S_{int})
\ra,
\label{pi1}\ee
where $(2\pi\b\hbar)^{-{D/2}}$ is the usual Feynman measure and
$\la x_f|$ and $|x_i\ra$ are both normalized as $\int d^D x \sqrt{g(x)}
|x\ra\la x| = 1$.
The factor $\l[ {g(x_f)\over g(x_i)} \r]^{1/4}$ in the measure
is a direct result from this normalization.\footnote{
Inserting $N$ sets of $p$-eigenstates and $N-1$ sets of $x$-eigenstates
gives a factor $\l[ {g(x_f) g(x_i)} \r]^{-{1/4}}$,
but the phase space path integration over the $p$'s and the
$q$'s gives an extra factor $g(x_f)^{1\over2}$ because there is one more
$p$ than $q$'s. This extra factor is taken at the point $x_f$ because we
 expanded the metric around this point. In the configuration space path
 integral,
 one integrates over the $p$'s, exponentiates the factors $g(x)$
by introducing ghosts, and then produces factors of $g(x)$
by integrating over the ghosts, with the same final result.}
Furthermore, $ \tau = {t\over \beta}$, and
$x^\mu(\tau) = x^\mu_{bg}(\tau)+q^\mu(\tau)$, where
$ x^\mu_{bg}(\tau) = x^\mu_f  + \tau(x^\mu_f- x^\mu_i)$ is the background
solution of the free equations which satisfies the correct boundary
conditions, and $q^\mu(\tau)$ are the quantum fluctuations.

The phase space interactions and propagators are given by
\eqa
\non  -{1\over \hbar}S_{int}^{phase}&=& -{1\over \beta\hbar }
\int^0_{-1}   d\tau \ {1\over 2}
\bigg ( g^{\mu\nu}(x) -  g^{\mu\nu}(x_f)\biggr ) \tilde p_\mu \tilde p_\nu
+ {i \over  \beta \hbar}
\int^0_{-1}   d\tau \ \tilde p_\mu (x^\mu_f - x^\mu_i)\\
&& - \beta \hbar \int_{-1}^{0} d\tau \ \Delta V_{W};
\label{phsp}
\eqae
\beq
\la \tilde p_\mu(\tau) \tilde p_\nu(\sigma)\ra
&=& \beta\hbar g_{\mu\nu}(x_f) \sim -1,
\nonumber\\
\la q^\mu(\tau) \tilde p_\nu(\sigma)\ra &=& -i \beta\hbar \delta^\mu_\nu
[\tau + \theta(\sigma -\tau)] \sim i \Delta^\bullet,
\label{phsp2}
\\
\la q^\mu(\tau)q^\nu(\sigma)\ra &=& -\beta\hbar g^{\mu\nu}(x_f)
\D(\tau,\sigma) \sim \Delta \nonumber,
\eeq
where we recall
\be
\D(\tau,\sigma)=\tau(\si+1)\te(\tau-\si)+\si(\tau+1)\te(\si-\tau).
\ee
We have rescaled $p_\mu(\sigma)$ into ${1 \over \beta} \tilde p_\mu(\sigma)$
in order that all propagators have a factor $\beta \hbar$
and all vertices a factor $(\beta \hbar)^{-1}$. The $\beta$ and
$\hbar$ appear always together and count the number of loops.
The configuration space interactions and propagators are as in 
section~\ref{II}, though regulated as in footnote~\ref{tmdsc}
\eqa
\non  -{1\over \hbar}S_{int}^{conf}&=& -{1\over \beta\hbar }
\int^0_{-1}   d\tau \ {1\over 2}
\bigg ( g_{\mu\nu}(x) -  g_{\mu\nu}(x_f)\biggr ) (\dot x^\mu \dot x^\nu +
a^\mu a^\nu + b^\mu c^\nu)  - \beta \hbar \int_{-1}^{0} d\tau\
\Delta V_{W};
\eqae
\beq
\la q^\mu(\tau) q^\nu(\si)\ra &=&
-\beta g^{\mu\nu}(x_f) \D(\tau,\si)
\sim \D
\nonumber\\
\la q^\mu(\tau) \dot q^\nu(\si)\ra &=& -\beta g^{\mu\nu}(x_f)
(\tau + \theta(\sigma-\tau))
\sim \D^{\bullet}
\nonumber\\
\la \dot q^\mu(\tau) \dot q^\nu(\si)\ra &=&
-\beta g^{\mu\nu}(x_f) (1 - \delta(\tau -\sigma))
\sim
{}^{\bullet\!}\D^{\bullet}
\label{config}
\\
\la a^\mu(\tau) a^\nu(\si)\ra &=&  \beta g^{\mu\nu}(x_f)
\delta(\tau-\sigma)
\sim -{}^{\bullet\bullet\!}\D
\nonumber\\
\la b^\mu(\tau) c^\nu(\si)\ra &=& -2\beta g^{\mu\nu}(x_f)
\delta(\tau-\sigma)
\sim 2 {}^{\bullet\bullet\!}\D
\nonumber
\eeq

\begin{figure}[htb]
$$
\begin{array}{rccc}
&{\mbox{\underline{Phase space}}} &&{ \mbox{\underline{Configuration space}}}
 \\[0.3in]
A:& \raisebox{-.1in}{\epsfxsize=0.5in \epsfbox{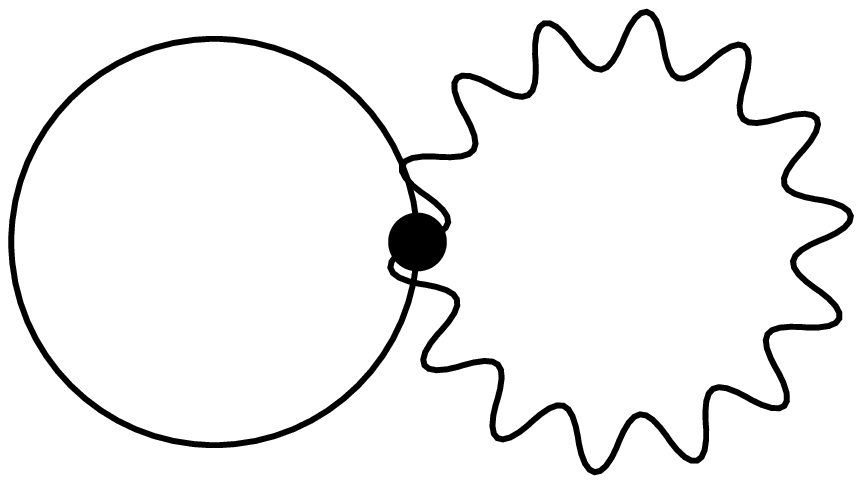}} &=& - \;\; 
\raisebox{-.2in}{\epsfxsize=0.5in \epsfbox{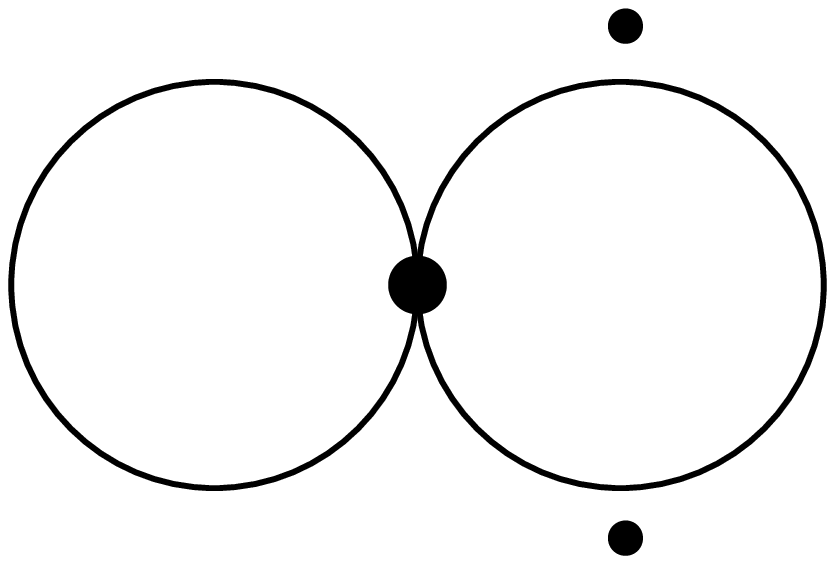}} \;\; -\;\; 
\raisebox{-.1in}{\epsfxsize=0.5in \epsfbox{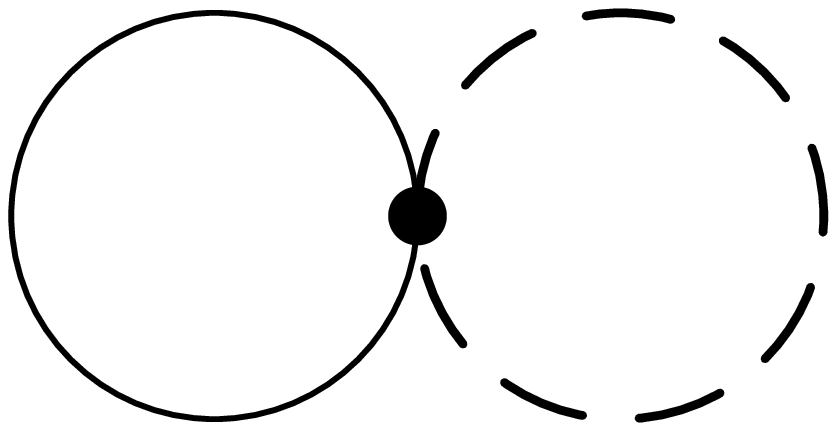}}  \\[0.15in]
B:& \raisebox{-.1in}{\epsfysize=0.3in \epsfbox{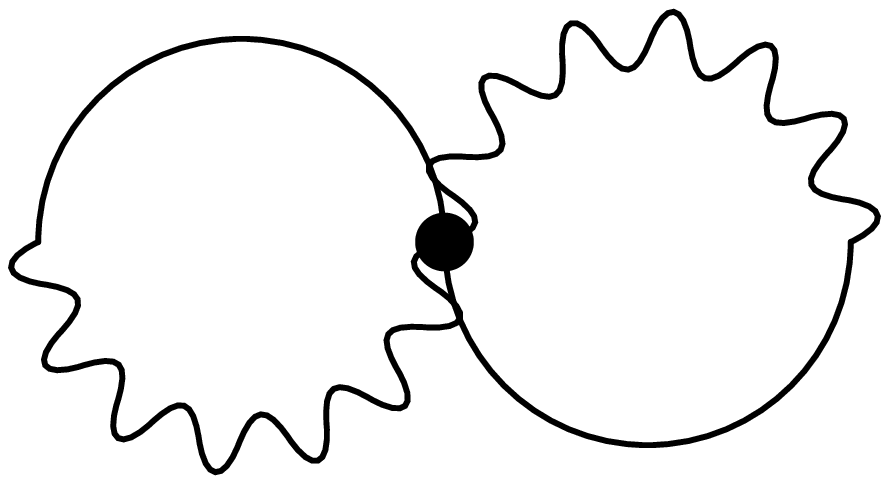}} &=& -
 \;\;\raisebox{-.2in}{\epsfxsize=0.5in \epsfbox{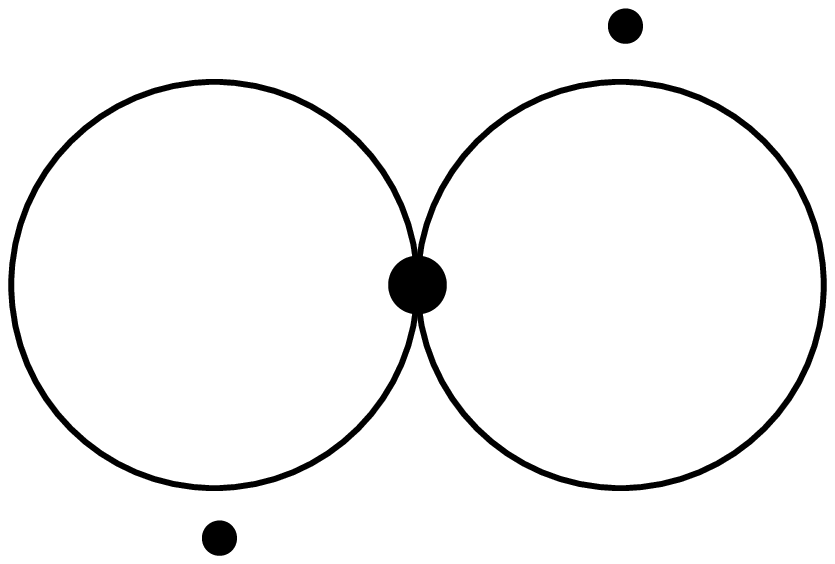}} \\[0.15in]
C:& \raisebox{-.1in}{\epsfysize=0.3in \epsfbox{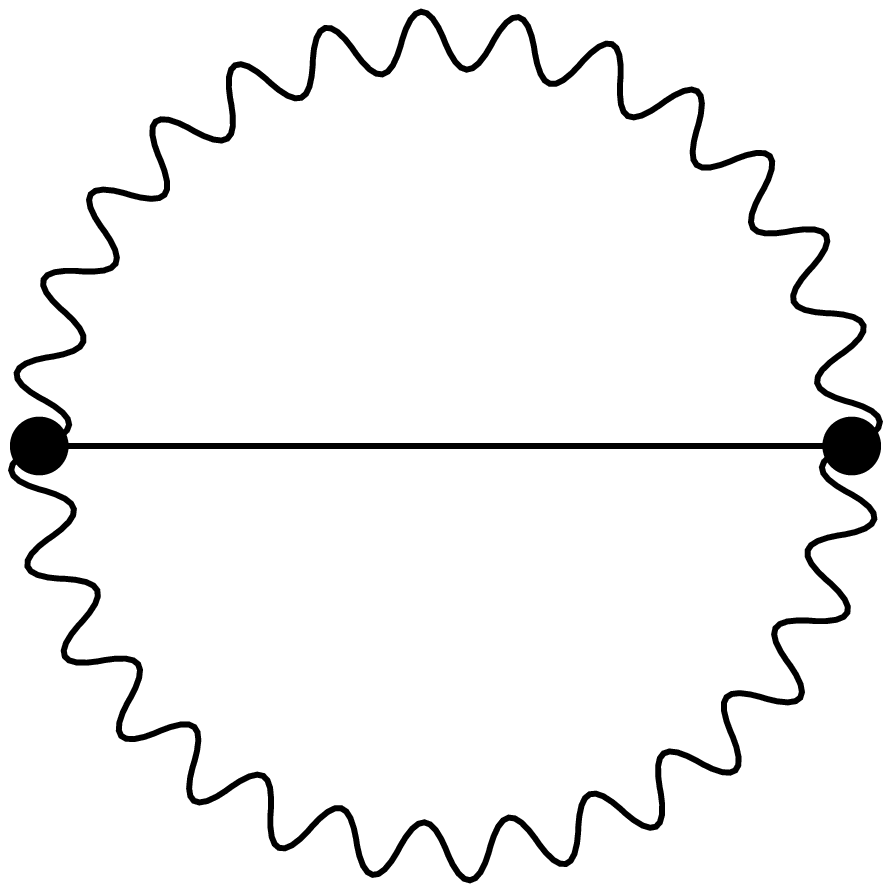}}\;\;-\,2 
(\beta\hbar)\;\;\raisebox{-.1in}{\epsfysize=0.3in \epsfbox{mod15.ps}} &=&
 \raisebox{-.1in}{\epsfysize=0.3in \epsfbox{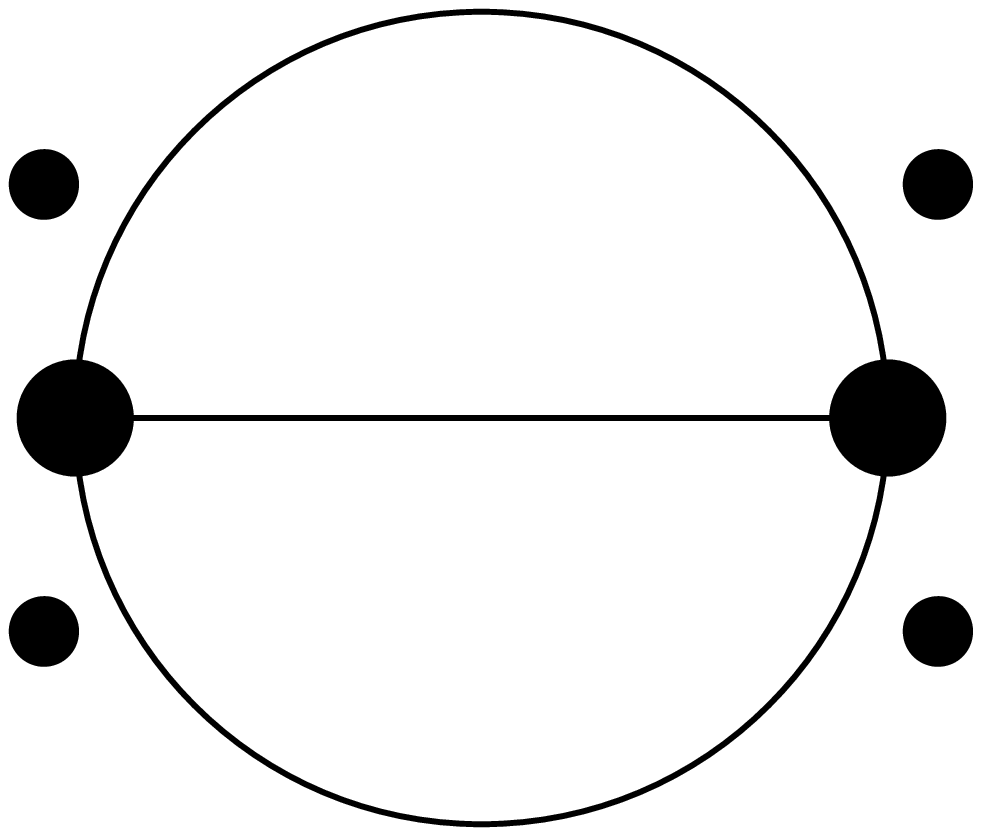}}\;\; - 
\;\;\raisebox{-.1in}{\epsfysize=0.3in \epsfbox{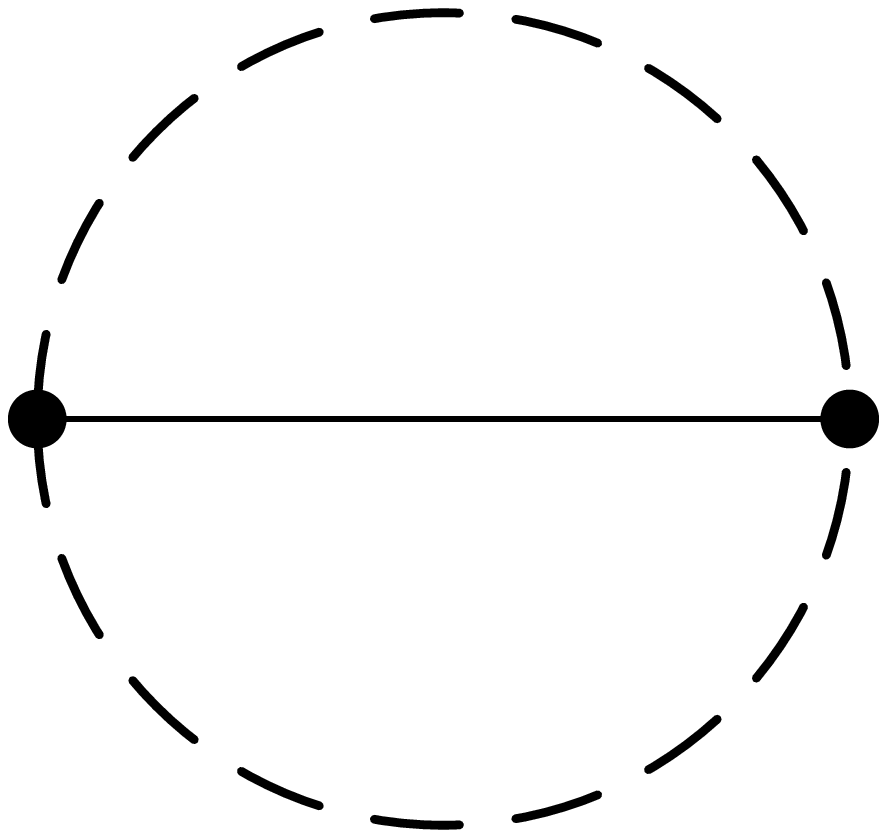}}\\[0.15in]
D:&  \raisebox{-.1in}{\epsfysize=0.3in \epsfbox{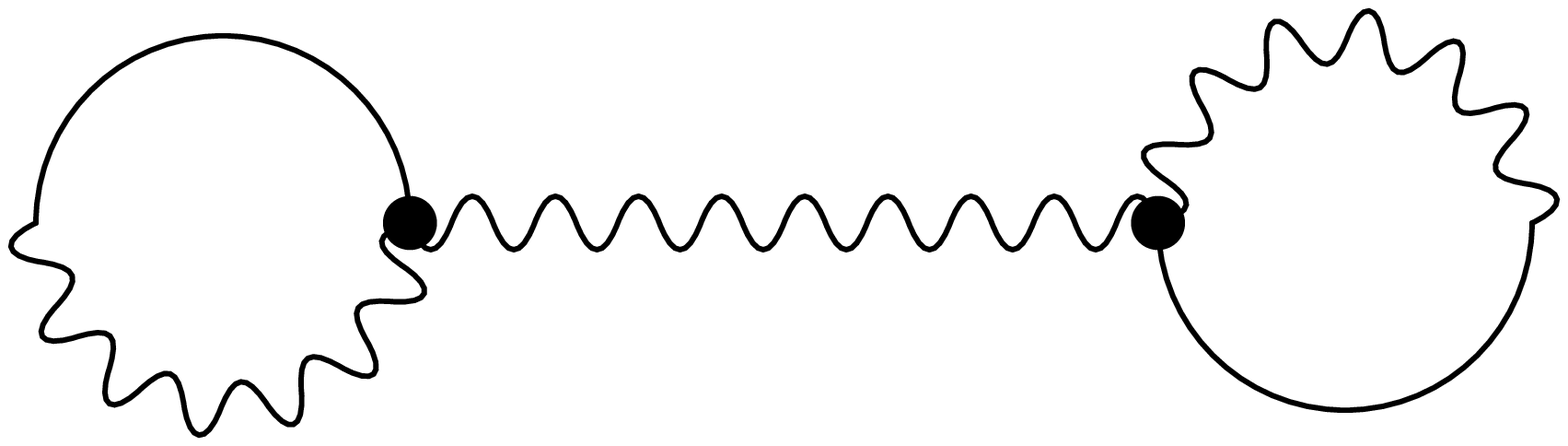}}\;\; 
-(\beta\hbar) \;\; \raisebox{-.1in}{\epsfysize=0.3in \epsfbox{mod19.ps}}&=&  
\raisebox{-.1in}{\epsfysize=0.3in \epsfbox{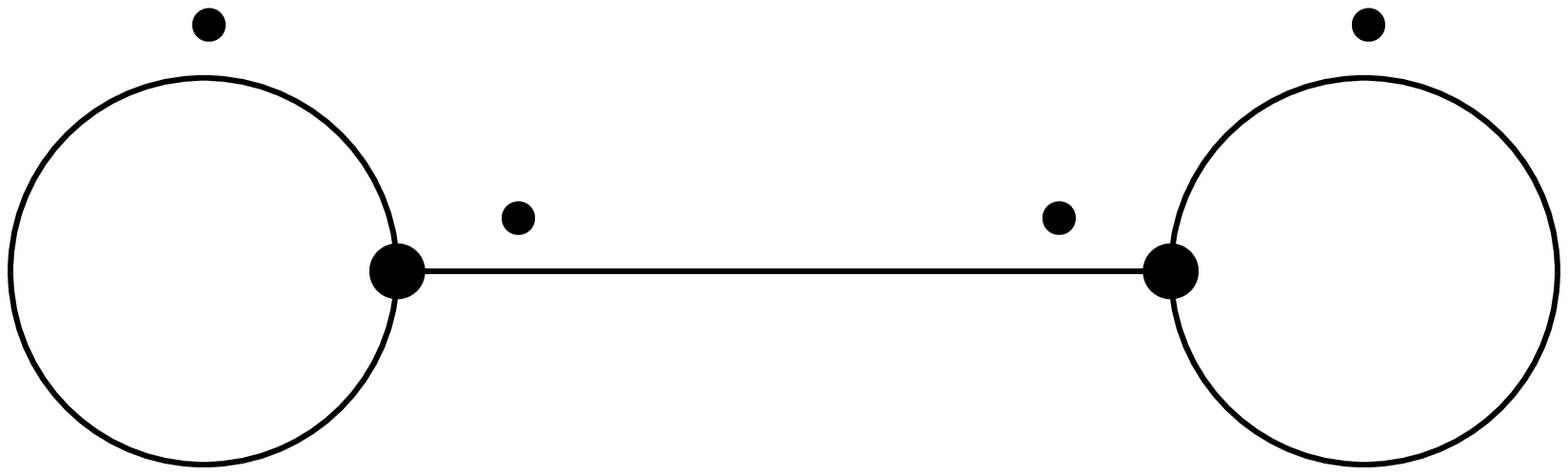}}\\[0.15in]
E:& \raisebox{-.1in}{\epsfysize=0.3in \epsfbox{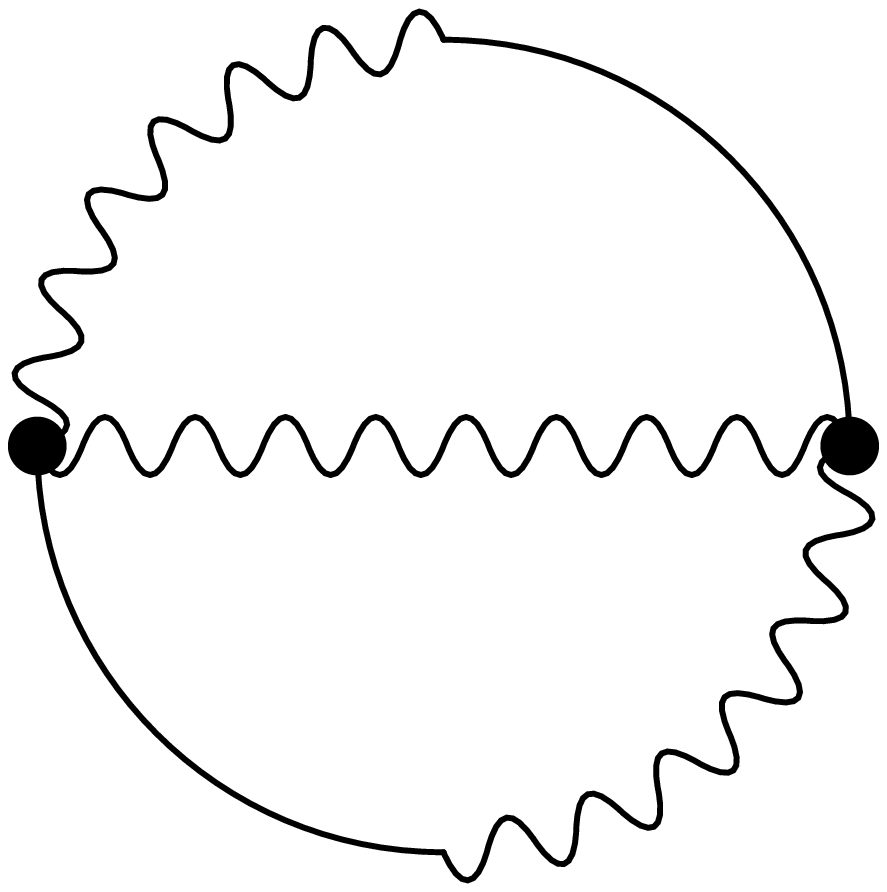}}\;\; -(\beta\hbar)\;\;
 \raisebox{-.1in}{\epsfysize=0.3in \epsfbox{mod19.ps}} &=& 
\raisebox{-.1in}{\epsfysize=0.3in \epsfbox{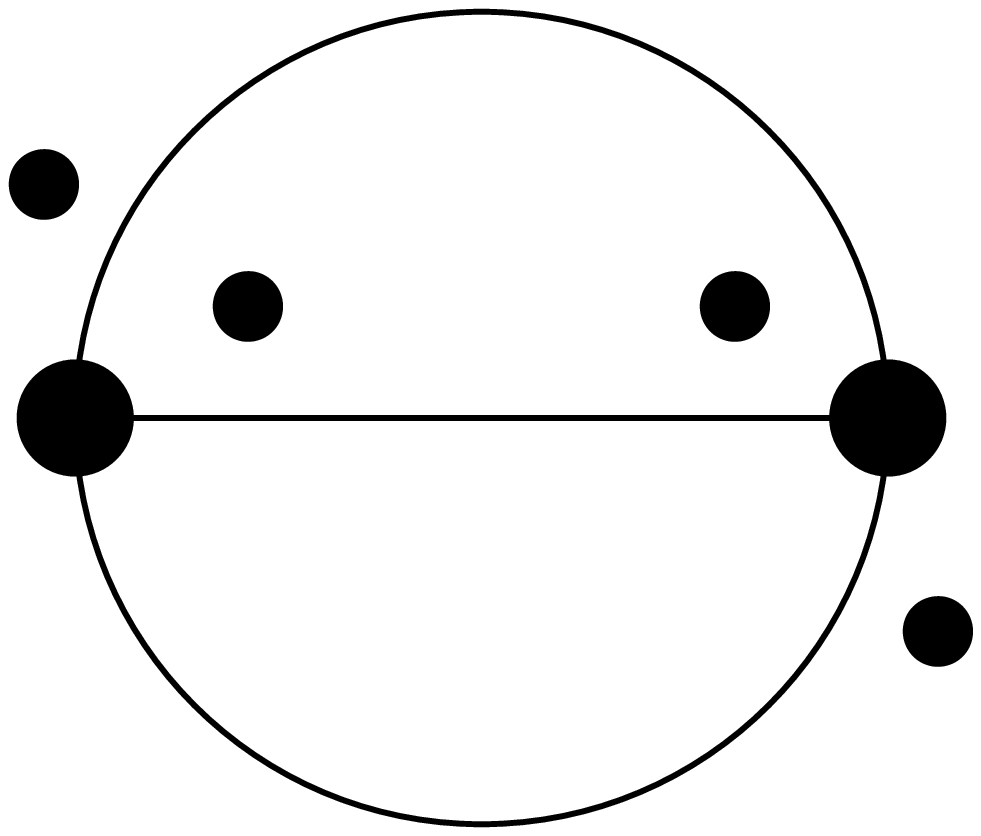}}\\[0.15in]
F:&\raisebox{-.1in}{\epsfysize=0.3in \epsfbox{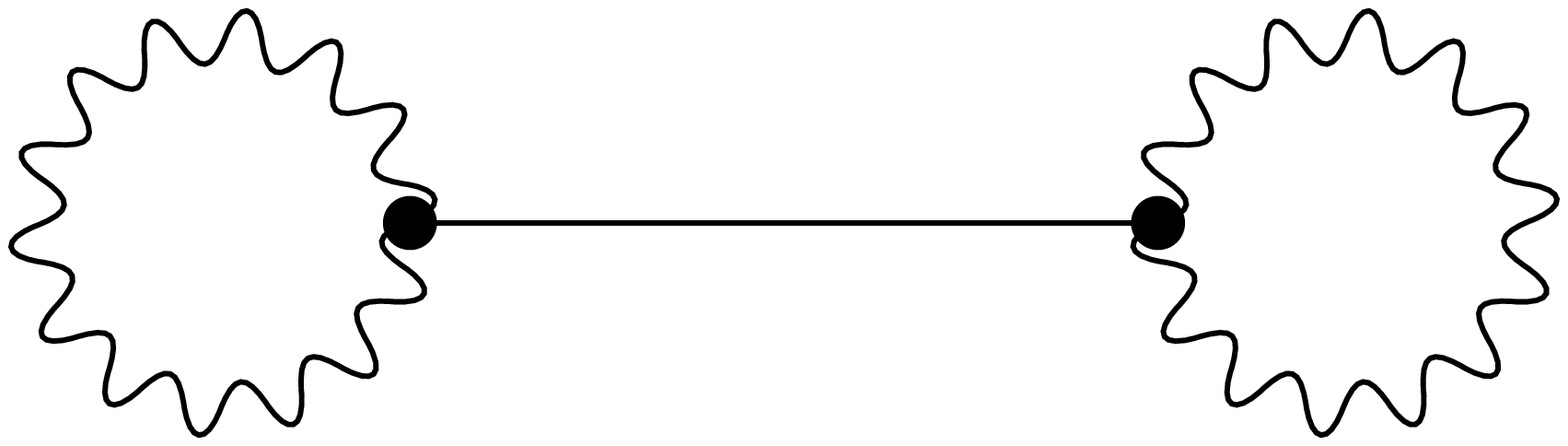}} &=& 
\left(\raisebox{-.22in}{\epsfxsize=0.5in \epsfbox{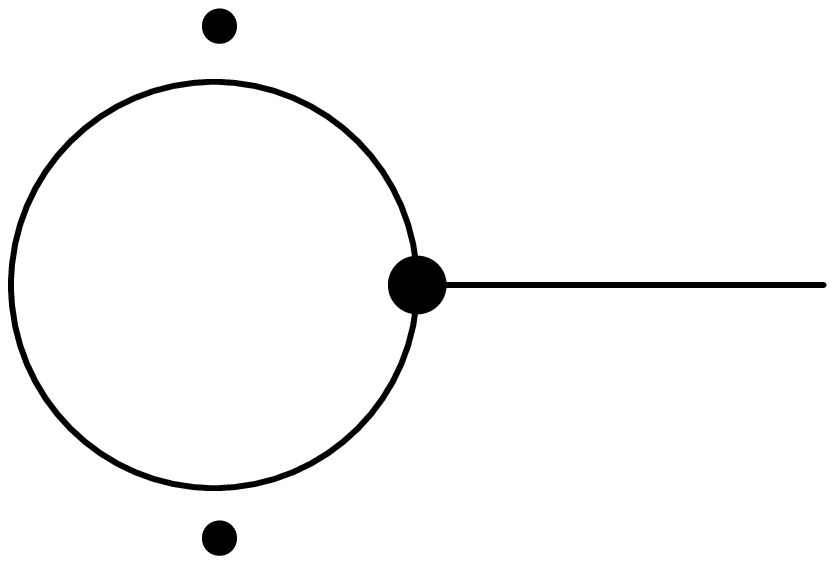}}+
\raisebox{-.1in}{\epsfxsize=0.5in \epsfbox{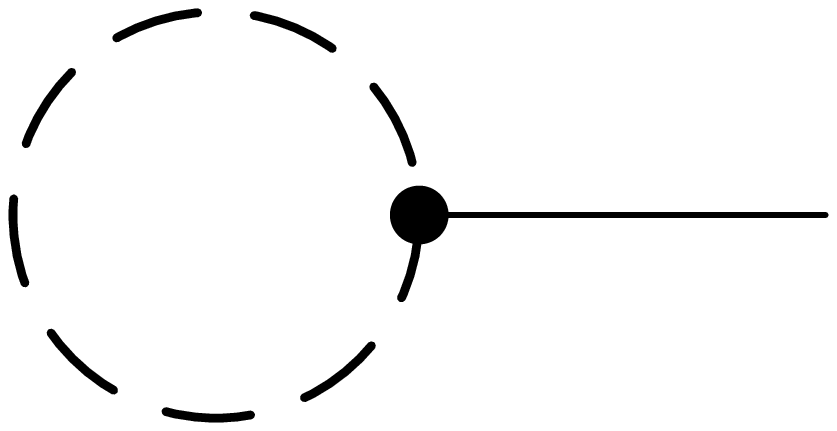}}\right)
\left(\raisebox{-.22in}{\epsfxsize=0.5in \epsfbox{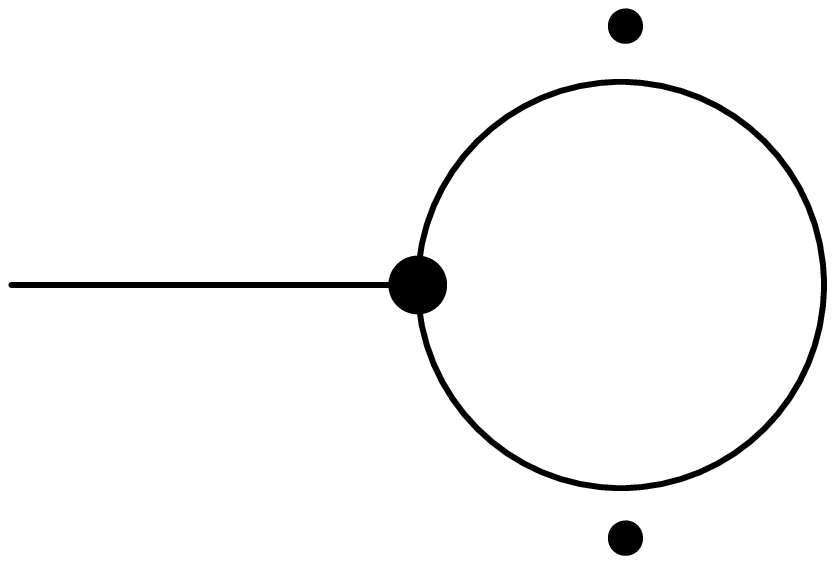}}+
\raisebox{-.1in}{\epsfxsize=0.5in \epsfbox{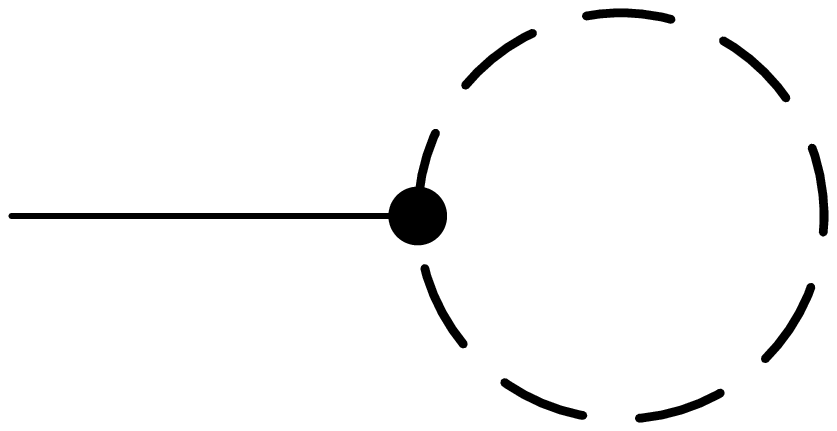}}\right)\\[0.15in]
G:&\raisebox{-.1in}{\epsfysize=0.3in \epsfbox{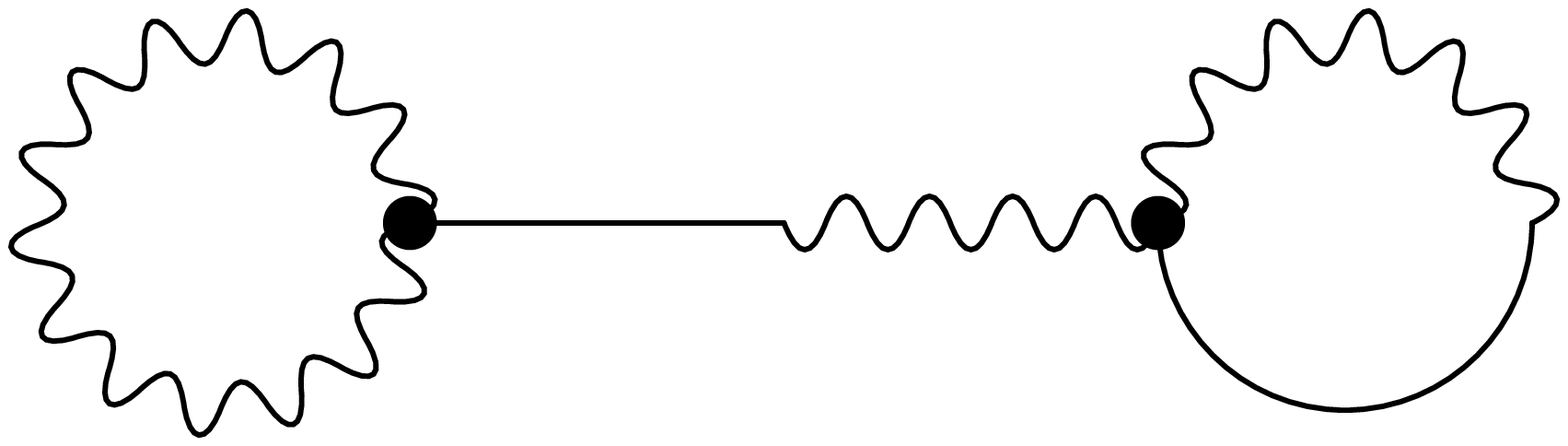}} &=& 
\left(\raisebox{-.22in}{\epsfxsize=0.5in \epsfbox{mod7.ps}}+
\raisebox{-.1in}{\epsfxsize=0.5in \epsfbox{mod9.ps}}\right)\left(
\raisebox{-.1in}{\epsfxsize=0.5in \epsfbox{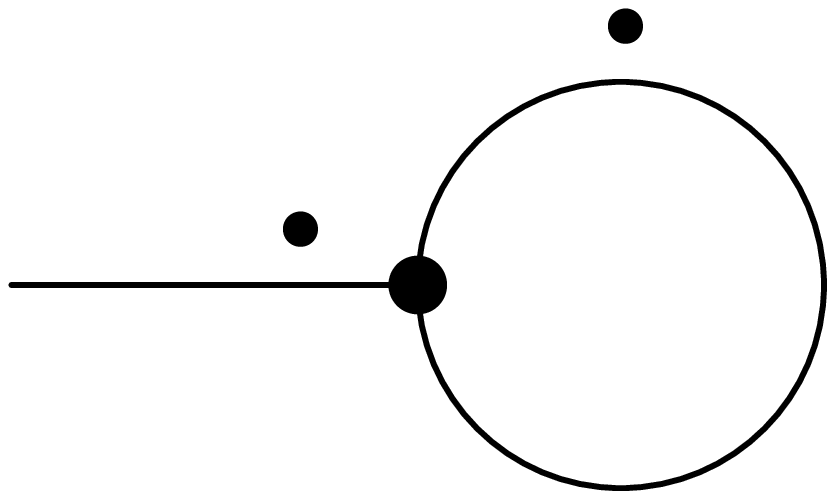}}\right)
\end{array}
$$
\caption{Two-loop diagrammatic identities using time slicing. Each
 cor\-res\-ponds to the coefficient of a particular metric structure present 
in the two-loop partition function; \mbox{$A$: $-\frac{1}{4}\,g^{\mu \nu}\Box
 g_{\mu\nu}$}, \mbox{$B$: $-\frac{1}{4}\,g^{\alpha \mu}g^{\beta\nu}
\pa_{\alpha}\pa_{\beta}g_{\mu\nu}$}, \mbox{$C$: $-\frac{1}{4}\,g^{\alpha\beta}
 g^{\mu\rho} g^{\nu \sigma} \pa_{\alpha}g_{\mu\nu}\pa_{\beta}g_{\rho\sigma}$},
 \mbox{\mbox{$D$: $-\frac{1}{2}\,g^{\alpha\mu} g^{\beta\rho} g^{\nu \sigma} 
\pa_{\alpha}g_{\mu\nu}\pa_{\beta}g_{\rho\sigma}$}, \mbox{$E$: 
$-\frac{1}{2}\,g^{\alpha\rho} g^{\beta\mu} g^{\nu \sigma} 
\pa_{\alpha}g_{\mu\nu}\pa_{\beta}g_{\rho\sigma}$}, 
\mbox{$F$: $-\frac{1}{8}\,g^{\alpha\beta} g^{\mu\nu} g^{\rho \sigma} 
\pa_{\alpha}g_{\mu\nu}\pa_{\beta}g_{\rho\sigma}$}}, and \mbox{$G$:
 $-\frac{1}{2}\,g^{\alpha\mu} g^{\beta\nu} g^{\rho \sigma} 
\pa_{\alpha}g_{\mu\nu}\pa_{\beta}g_{\rho\sigma}$}.}
\label{bella}
\end{figure}

From $ {}^\bullet\! \Delta^\bullet = 1-\delta (\tau-\sigma)$
and ${}^{\bullet\bullet} \Delta = \delta (\tau-\sigma)$
it is clear that the sum of the $\la \dot{q} \dot{q} \ra$ and the ghost
 propagators is equal to minus
the $\la pp\ra $ propagator.
\be
\raisebox{.01in}{\epsfxsize=0.6in
	\epsfbox{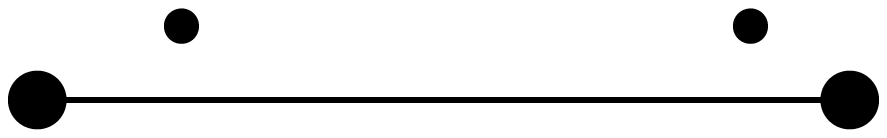}} \;\;+ \;\;\raisebox{.01in}{\epsfxsize=0.6in
	\epsfbox{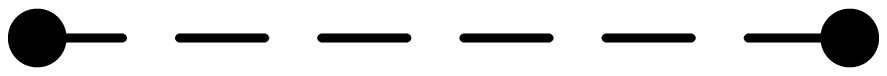}}\;\; =\;\; - \left(
\raisebox{.01in}{\epsfxsize=0.6in
\epsfbox{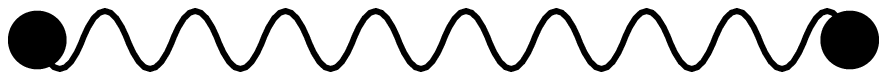}}\right).
\label{chick}
\ee
 Further, the $\la q p\ra$ and the $\la q
\dot q\ra$
propagators are  equal up to a factor $i$ (note that $p=i\dot x$ is a field
equation).
\be
\raisebox{.01in}{\epsfxsize=0.6in
	\epsfbox{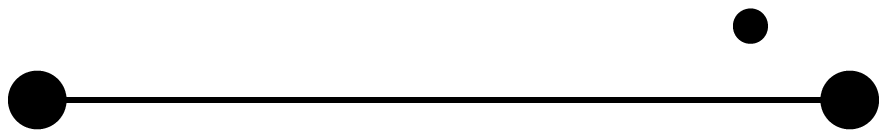}} \;\;\; \times i \;\;\;=\;\;\; 
\raisebox{.01in}{\epsfxsize=0.6in
\epsfbox{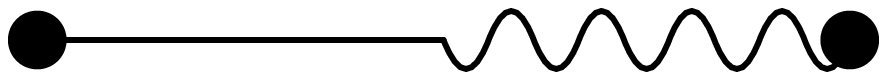}}
\label{pox}
\ee
We are now ready to evaluate two-loop diagrams. The result for the 
\mbox{$\xi^{\mu}=x_i^{\mu}-x_f^{\mu}$} independent part is given in
 figure~\ref{bella} In the phase space approach, one finds derivatives of 
the inverse
metric $g^{\mu\nu}$ which we have converted to derivatives of the metric
$g_{\mu\nu}$ for purpose of comparison. The relations $A,\,B,\,F$, and $G$ 
follow immediately from~(\ref{chick}) and~(\ref{pox}). Considering the 
identities $C$, $D$ and $E$ one must note that the phase space expression 
comes from both one- and two-vertex graphs. The identities then follow from
 rewriting~(\ref{chick}) as
$$
\raisebox{.01in}{\epsfxsize=0.6in
	\epsfbox{mod21.ps}} \;\;=\;\; - 
\;\;\raisebox{.01in}{\epsfxsize=0.6in
\epsfbox{mod22.ps}} \;\; - \;\;\raisebox{.01in}{\epsfxsize=0.6in
	\epsfbox{mod23.ps}}\;\; 
$$
and replacing a ``ghost'' propagator on the phase space side by its value, 
$(-\beta\hbar)$ times a delta-function, which pinches the two vertices. It is
 clear that the two-loop corrections in configuration space agree with those 
in phase space. We have also checked the $\xi^{\mu}$ dependent terms in the 
transition element and found complete agreement with~(\ref{pex}). As, once 
again, all configuration space graphs at the three- and higher loop level are 
unambiguous, the two-loop calculation suffices to show the equivalence between
 the phase- and configuration-space path integrals.  

\section{Conclusions}
\label{IV}
\hspace{.5in}

The two regularization schemes considered in this article, mode regularization
 and time slicing, give the same answers at all orders of loops provided one 
adds specific order $\hbar^2$ counterterms to the action in the
 configuration space path integral. These terms are proportional to the 
curvature $R$ but also to products of two Christoffel symbols and the latter 
are different for the two schemes. These extra terms of order 
 $\hbar^2$ follow in the time slicing method from rewriting the Hamiltonian 
in Weyl-ordered form, but for mode regularization we could only determine 
them by requiring that the Schroedinger equation be satisfied. (This way of
 fixing the extra terms can also be applied to time slicing). As a nontrivial
 check that these extra terms are indeed needed (and also a check on their 
explicit form) we evaluated the three-loop contributions to the trace anomaly 
in both schemes and indeed found the same (correct) answer. 

We also considered phase space path integrals. Their loop graphs contain no 
divergences. We gave a diagrammatic proof of Matthews' theorem that phase and
 configuration space path integrals give the same transition element. A key
 observation was that the difference between the phase space propagator 
$\la p(\sigma) p(\tau) \ra$ and the configuration space  propagator
 $\la \dot{q}(\sigma) \dot{q}(\tau) \ra$ is equal to the propagator for the 
Lee-Yang ghosts. The latter result when one exponentiates the factors 
$\sqrt{g(x)}$ which are produced by integrating out the momenta.

It remains to explain why the coefficient of the $R$ term in $\Delta
 V$ is equal to $\frac{1}{8}$ in all ``reasonable'' regularization
 schemes. Let us  call a scheme reasonable if it satisfies Matthews'
 theorem. Then the first  two graphs in figure~1 should give the same
 result in configuration space as  in phase space. The equality for
 graph $B$ is always satisfied, see~(\ref{phsp2}) and~(\ref{config}), but for
 graph $A$ one finds the condition 
\be \int_{-1}^0 d\tau
 \:\Delta(\tau,\tau)=\int_{-1}^0 d\tau
 \left[\Delta(\tau,\tau)\left.\biggl(
 {}^{\bullet}\Delta^{\bullet}(\tau,\sigma) +
 {}^{\bullet\bullet}\Delta(\tau,\sigma)\biggr)\right|_{\sigma=\tau}\right].
 \ee Both time slicing and mode regularization satisfy this condition.
 In all such reasonable schemes the transition element with the naive
 action  $S=\frac{1}{2} g_{\mu\nu}(x) \dot{x}^{\mu}\dot{x}^{\nu}$ will
 then contain a  two-loop contribution $\frac{\hbar^2}{24}R$ just as
 the phase space calculation would give.  The complete phase space
 Hamiltonian~(\ref{phsp}) contains, however,  in addition the
 counterterm~(\ref{eggs}). The total $R$ term in the Hamiltonian element
is then $-\frac{\hbar^2}{12} R$ for any reasonable regularization scheme and 
the $R$ term in the  action to be used in the
 configuration space path integral has  coefficient $\frac{1}{8}$.

\end{document}